# Closed-Form EGN Model with Comprehensive Raman Support

Yanchao Jiang[(1)], Antonino Nespola[(2)], Stefano Straullu[(2)], Alberto Tanzi[(3)],
Stefano Piciaccia[(3)], Fabrizio Forghieri[(3)], Dario Pilori[(1)], Pierluigi Poggiolini[(1)]

[(1)] DET, Politecnico di Torino, C.so Duca Abruzzi 24, 10129, Torino, Italy, yanchao.jiang@polito.it
[(2)] LINKS Foundation, Via Pier Carlo Boggio 61, 10138, Torino, Italy
[(3)] CISCO Photonics Italy srl, via Santa Maria Molgora 48/C, 20871, Vimercate (MB), Italy

**Abstract** *We present a series of experiments testing the accuracy of a new closed-form multiband EGN model, carried out over a full-Raman 9-span C+L link. Transmission regimes ranged from linear to strongly non-linear with large ISRS. We found good correspondence between predicted and measured performance.* ©2024 The Author(s)

**Introduction**
In optical networking, it has become crucial to incorporate the physical-layer behavior into the network design, management, control and optimization processes. To this end, physical-layer models (PLMs) have been developed that have achieved significant success in accurately predicting optical link behavior. These PLMs have been extensively utilized in the industry for some time, with the GN and EGN models being among the most widely adopted.

Recently, though, the requirements on PLMs have been stepped up significantly. They now ask for faster computation speeds for iterative optimization and real-time management, as well as the ability to account for frequency-dependent parameters, due to the extension of the usable fiber bandwidth to multiband (MB) [1], [2].

To address these challenges, approximate closed-form models (CFMs) have been developed, capable of real-time computation, supporting MB by including the frequency-dependence of all key parameters, as well as Inter-channel Raman scattering (ISRS). Two groups, one at UCL and one at PoliTo (in collaboration with CISCO) have independently obtained MB-enabled CFMs that are based on the GN and EGN models, with similar foundations and capabilities but with differences in features and final analytical form. For the CFM by UCL see [3] (and [4]-[8]). For the CFM by PoliTo and CISCO, which we call CFM5, see [9] (and [10]-[15]). In [15] we recently carried out a thorough validation campaign of CFM5, consisting of several experiments, performed on a 5-span, 430km C+L link.

CFM5 is capable of accurately modelling MB systems with ISRS and forward-pumped Raman amplification but cannot account for non-linear interference (NLI) that is produced at the end of the span in the presence of *backward-pumped* Raman amplification (BPRA). Since BPRA could potentially be quite important in MB systems [16], we developed a new model (CFM6, see details in [17]) that can account for NLI due to BPRA too. In this paper we first briefly introduce the model, then we report on a 9-span, 540km C+L experiment which uses BPRA *only*. A variety of propagation conditions were tested, ranging from linear to very deeply non-linear. In some of them, NLI from BPRA was the single strongest contribution to overall noise. In all testing conditions CFM6 showed quite good agreement with the experiments.

**The closed-form model CFM6**
CFM5 was shown as Eqs. (1)-(6) in [9]. CFM6 is introduced below, using the same formula numbers. The field loss (units 1/km) in any given span, is described as:

$$\alpha_{n,i}(z) = \alpha_{0,n,i} + \alpha_{1,n,i} \cdot \exp(-\sigma_{n,i} \cdot z) \quad (1)$$

where: $n$ is the channel index, identifying the channel located at frequency $f_n$; $i$=1 is for loss profile in the forward direction, with z=0 at the span start; $i$=2 for loss profile in the backward direction, with z=0 at the span end. CFM5 used only $i$=1. To assign the 6 parameters per channel required by (1), we first numerically solve the Raman differential equations (Eq. (1) in [14]) and find the *exact* power evolution of every channel. Then Eqs. (30.1) and (30.2) in [13] provide the values of $\alpha_{0,n,i}$ and $\alpha_{1,n,i}$ which minimize the Mean Square Error (MSE) between the power profile generated by (1) and the exact power evolution, *for a given value* of $\sigma_{n,i}$. Numerical optimization over $\sigma_{n,i}$ finds the final values of $\alpha_{0,n,i}, \alpha_{1,n,i}, \sigma_{n,i}$.

Then, the NLI power $P_n^{\text{NLI}}$ generated in a given span, affecting the $n$-th channel, can be found using Eq.(2), where $P_{m,i}$, $R_m$ and $\bar{\beta}_{2,m}$ are for the $m$-th channel respectively: the launch power into the span ($i$=1) and out of the span ($i$=2), the symbol rate and the 'effective dispersion'. The latter is given by:

$$\bar{\beta}_{2,m} = \beta_2 + \pi\beta_3(f_n + f_m - 2f_0) + \frac{2\pi^2}{3} \cdot \\ \cdot \beta_4[(f_n - f_0)^2 + (f_n - f_0)(f_m - f_0) + (f_m - f_0)^2] \quad (3)$$

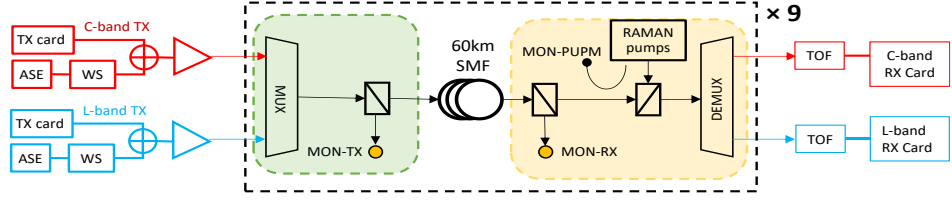

**Fig. 1:** General schematic of the C+L Raman-only-amplified line experiment.
TOF: 50-GHz bandwidth tunable optical filter. WS: WaveShaper™ (configurable optical filter).

$$P_n^{\text{NLI}} = \frac{16}{27} \sum_{\substack{1\le m\le N_c,\, 0\le j\le 1 \\ 0\le k\le Q_i,\, 0\le q\le Q_i \\ 1\le i\le 2}} \frac{\gamma_{n,m}^2 \cdot P_{n,i} \cdot P_{m,i}^2 \cdot \rho_m \cdot (2-\delta_{m,n}) \cdot e^{-4\alpha_{1,m,i}/\sigma_{m,i}} \cdot (-1)^j}{2\pi R_m^2 \cdot k!\, q!\, \bar{\beta}_{2,m} \cdot (4\alpha_{0,m,i} + (k+q)\sigma_{m,i})} \cdot \left(\frac{2\alpha_{1,m,i}}{\sigma_{m,i}}\right)^{k+q} \Psi_{m,n,j,k,i} \quad (2)$$

where $f_0$ is the frequency where $\beta_2$, $\beta_3$ and $\beta_4$ are measured. The non-linearity coefficient $\gamma_{n,m}$ has the frequency-dependent expression:

$$\gamma_{n,m} = \frac{2\pi f_n}{c} \frac{2n_2}{A_{\text{eff}}(f_n)+A_{\text{eff}}(f_m)} \quad (4)$$

Furthermore, in Eq. (2):

$$\Psi_{m,n,j,k,i} = \operatorname{asinh}\left(\frac{\pi^2 \bar{\beta}_{2,m} R_n \cdot (f_m - f_n + (-1)^j \cdot R_m/2)}{2\alpha_{0,m,i} + k\sigma_{m,i}}\right)(5)$$

Also, in Eq. (2), $\delta_{m,n}$ is 1 if $m=n$ and 0 otherwise; $N_c$ is the number of WDM channels; $Q_i$ represents the order of a series expansion, whereby more terms are called into play as the strength of Raman requires, and is set by [9]:

$$Q_i = \max(\lfloor 10 \cdot |2\alpha_{1,n,i}/\sigma_{n,i}| \rfloor + 1) \quad (6)$$

The max in (6) is taken across all channels ($1 \le n \le N_c$) in the WDM comb. Finally, $\rho_m$ is the machine-learning-based 'correction term' given by Eq. (14) in [12] (Table IV). Note that $\rho_m$ depends on the span index, like most other quantities in Eq. (2). However, to avoid clutter, we have omitted to indicate such dependence. $\rho_m$ improves the accuracy of the CFM, as shown in [12], allowing it to closely approach the EGN model. $\rho_m$ depends in closed-form on $R_m$, on the accumulated dispersion and on the modulation format of the $m$-th channel (see [12] for details).

Eq. (2) provides the $P_n^{\text{NLI}}$ generated on the $n$-th channel within *one span* of the link. To obtain the total NLI at the end of the link it is enough to sum the $P_n^{\text{NLI}}$ of each span. The total NLI can then be used together with ASE noise to find the generalized SNR (or GSNR) for the $n$-th channel at the end of the link.

**The experiment setup**
The schematic is shown in Fig.1. The WDM comb was generated by shaping ASE noise through programmable optical filters, emulating 32 GBaud channels spaced 50 GHz, roll-off 0.1. The reason for choosing a low Baud rate will be clarified later. The WDM comb consisted of a variable number of channels in C-band and L-band (see details later, Fig.4). For measurement, each emulated channel was replaced in turn by an actual PM-32QAM channel, generated by either a C-band or a L-band transmitter card.

The line consisted of 9 spans of SMF G652D each of about 60km uninterrupted length (no connectors within the span). At each span end, five Raman pumps were injected in the backward direction, with the following frequencies and average power [THz, mW]: 210.6, 288.2; 208.9, 263.4; 206.7, 167.9; 204.5, 112.4; 200.6, 161.5. Pump powers differed slightly from span to span, so we report the average value.

The line was instrumented so that the WDM signal spectrum transmitted and received was carefully measured for each span (at "MON-Tx" and "MON-Rx"). The receivers were separate C and L-band units and provided hard-decision BER as well as the constellation SNR after DSP.

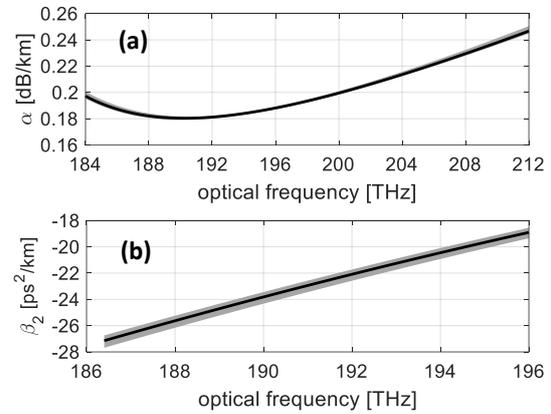

**Fig. 2:** measured (a) fiber attenuation and (b) dispersion, vs. optical frequency. Solid line: average value. Gray band: showing the peak-to-peak range among the 9 fiber measurements.

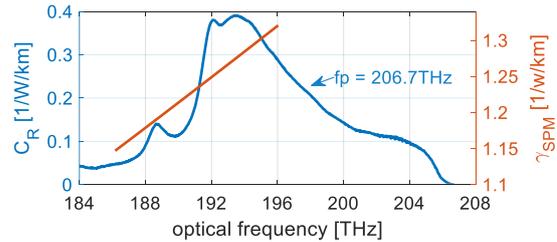

**Fig. 3:** Blue: Raman gain spectrum for the third pump at 206.7 THz. Orange: SPM non-linearity coefficient $\gamma_{n,n}$.

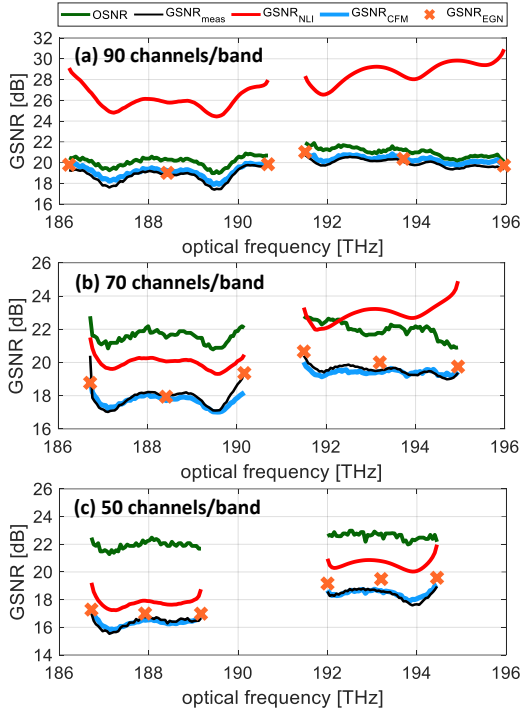

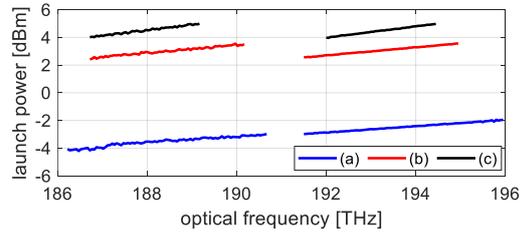

**Fig. 4:** Experiment results with increasing amount of non-linearity. Green line: measured OSNR at Rx (ASE only). Red line: GSNR due to NLI only, from CFM6. Black line: measured total GSNR (ASE+NLI). Blue line: total GSNR with NLI from CFM6. Black markers: total GSNR with NLI from numerically integrated EGN model.

**Fig. 5:** Launch power into the first span for (a)-(c), see Fig.4.

The 9 spans were characterized as for their attenuation (start and end connector loss removed) and dispersion profiles vs. frequency (Fig.2). The non-linearity coefficient $\gamma$ was characterized by means of a dedicated experiment based on XPM and found to agree with the literature-reported values of 1.25 1/(W km) at 192THz. As for the frequency variability of $\gamma$, the behavior described by Eq. (4) was assumed. Note that the equation provides different values depending on NLI being SPM ($n = m$) or XPM ($n \neq m$). A plot of $\gamma_{n,n}$ (SPM) vs. frequency is shown in Fig.3 as an example. Regarding ISRS, the Raman gain spectrum was measured at 206.7 THz (pump3, the blue curve in Fig.3) and then translated to different pump frequencies according to Eqs. (37)-(39) in [18].

**Results and conclusion**
The goal of this study was to validate CFM6 by comparing the GSNR predicted using CFM6 ($\text{GSNR}_{\text{CFM}}$, blue line) with the GSNR measured on all channels ($\text{GSNR}_{\text{meas}}$, black line), in different propagation regimes, see Fig.4.

$\text{GSNR}_{\text{meas}}$ was found as follows. The Rx provided $\text{SNR}_{\text{meas}}$, i.e., the constellation SNR after DSP. $\text{GSNR}_{\text{meas}}$ is then calculated as:
$\text{GSNR}_{\text{meas}} = \text{SNR}_{\text{meas}} \cdot (1 - \text{SNR}_{\text{meas}}/\text{SNR}_{\text{bb}})^{-1}$
where $\text{SNR}_{\text{bb}}$ is the constellation SNR measured in back-to-back (no ASE), which accounts for the internal Tx-Rx noise. We measured $\text{SNR}_{\text{bb}}$=22 dB for both the C- and L-band Tx-Rx pairs. The reason why we chose 32 GBaud symbol rate was because of the high value of $\text{SNR}_{\text{bb}}$. At 60 GBaud the value was down to 20dB, which would have negatively affected the ability of our tests to validate CFM6.

$\text{GSNR}_{\text{CFM}}$ was calculated using the NLI predicted by CFM6, based on the detailed characterization of the overall link and on the signal launch power spectrum into each span measured at "MON_TX" (Fig.1). The ASE noise in $\text{GSNR}_{\text{CFM}}$ was measured for each channel at the Rx. Fig.4 shows the results of the comparison among $\text{GSNR}_{\text{meas}}$ and $\text{GSNR}_{\text{CFM}}$ in different regimes (a)-(c). We also show GSNR found using the numerically integrated EGN model, as markers. The launch power into the first span is shown in Fig.5. For (a) propagation was essentially linear and indeed $\text{GSNR}_{\text{CFM}}$ and OSNR almost coincide. In (b) it was non-linear, especially in L-band. In (c) it was deeply non-linear. Note that we used fewer channels in (b), 70, and in (c), 50, vs. (a), 90. This was done to ensure a higher power per channel in (b) and (c) at launch and at each following span. Importantly, in all tests, the NLI produced at the end of each span was comparable or greater than that produced at the start, creating an ideal condition for testing CFM6, whose specific difference vs. CFM5 is the ability to account for such NLI. We had $P_{\text{NLI}}^{\text{end}}/P_{\text{NLI}}$: (a) mean 40%, max 49%; (b) mean 47%, max 62%; (c) mean 47%, max 54%. Overall, Fig.4 shows a good match between $\text{GSNR}_{\text{meas}}$ and $\text{GSNR}_{\text{CFM}}$. This is particularly significant especially in (b) and (c), where noise at the Rx is mostly NLI and $\text{GSNR}_{\text{CFM}}$ is mostly set by $P_{\text{NLI}}$ from CFM6.

In conclusion, we presented a closed-form approximate multiband EGN model (CFM6) with comprehensive Raman support. We tested it in a challenging all-Raman amplified C+L system experiment, which we designed to stress-test the CFM and check its ability to account for NLI produced at the end of each span. The results indicate good accuracy of the model.

**Acknowledgements**
Systems through the RISE-ONE Sponsored Research Agreement (SRA); the PhotoNext Center of Politecnico di Torino; the European Union under the Italian National Recovery and Resilience Plan (NRRP) of NextGenerationEU, partnership on "Telecommunications of the Future" (PE00000001 - program "RESTART").

**References**

[1] J. Renaudier *et al.*, "Devices and Fibers for Ultrawideband Optical Communications," *Proc. IEEE*, vol. 110, no. 11, pp. 1742–1759, 2022, doi: 10.1109/JPROC.2022.3203215.

[2] T. Hoshida *et al.*, "Ultrawideband Systems and Networks: Beyond C + L-Band," *Proc. IEEE*, vol. 110, no. 11, pp. 1725–1741, 2022, doi: 10.1109/JPROC.2022.3202103.

[3] H. Buglia *et al.*, "A Closed-Form Expression for the Gaussian Noise Model in the Presence of Inter-Channel Stimulated Raman Scattering Extended for Arbitrary Loss and Fibre Length," *J. Light. Technol.*, pp. 1–10, 2023, doi: 10.1109/JLT.2023.3256185.

[4] D. Semrau, G. Saavedra, D. Lavery, R. I. Killey, and P. Bayvel, "A Closed-Form Expression to Evaluate Nonlinear Interference in Raman-Amplified Links," *J. Light. Technol.*, vol. 35, no. 19, pp. 4316–4328, Oct. 2017, doi: 10.1109/JLT.2017.2741439.

[5] D. Semrau, R. I. Killey, and P. Bayvel, "A Closed-Form Approximation of the Gaussian Noise Model in the Presence of Inter-Channel Stimulated Raman Scattering." Aug. 23, 2018. [Online]. Available: https://arxiv.org/abs/1808.07940

[6] D. Semrau, R. I. Killey, and P. Bayvel, "A Closed-Form Approximation of the Gaussian Noise Model in the Presence of Inter-Channel Stimulated Raman Scattering," *J. Light. Technol.*, vol. 37, no. 9, pp. 1924–1936, May 2019, doi: 10.1109/JLT.2019.2895237.

[7] D. Semrau, E. Sillekens, R. I. Killey, and P. Bayvel, "A Modulation Format Correction Formula for the Gaussian Noise Model in the Presence of Inter-Channel Stimulated Raman Scattering," *J. Light. Technol.*, vol. 37, no. 19, pp. 5122–5131, Oct. 2019, doi: 10.1109/JLT.2019.2929461.

[8] D. Semrau, "Modeling of Fiber Nonlinearity in Wideband Transmission," in *OFC 2022*, San Diego (CA), Mar. 2022, p. paper W3C.6.

[9] P. Poggiolini and M. Ranjbar-Zefreh, "Closed Form Expressions of the Nonlinear Interference for UWB Systems," in *2022 European Conference on Optical Communication (ECOC)*, 2022, p. Tu1D.1.

[10] P. Poggiolini, G. Bosco, A. Carena, V. Curri, Y. Jiang, and F. Forghieri, "The GN model of fiber non-linear propagation and its applications," *J Light. Technol*, vol. 32, no. 4, pp. 694–721, Feb. 2014.

[11] P. Poggiolini, "A Generalized GN-Model Closed-Form Formula." Sep. 24, 2018. [Online]. Available: https://arxiv.org/abs/1810.06545v2

[12] M. Ranjbar Zefreh, F. Forghieri, S. Piciaccia, and P. Poggiolini, "Accurate Closed-Form Real-Time EGN Model Formula Leveraging Machine-Learning Over 8500 Thoroughly Randomized Full C-Band Systems," *J. Light. Technol.*, vol. 38, no. 18, pp. 4987–4999, Sep. 2020, doi: 10.1109/JLT.2020.2997395.

[13] M. Ranjbar Zefreh and P. Poggiolini, "A Real-Time Closed-Form Model for Nonlinearity Modeling in Ultra-Wide-Band Optical Fiber Links Accounting for Inter-channel Stimulated Raman Scattering and Co-Propagating Raman Amplification," *ArXiv Prepr. ArXiv200603088*, Jun. 2020, doi: 10.48550/arXiv.2006.03088.

[14] M. Ranjbar Zefreh and P. Poggiolini, "Characterization of the Link Function in GN and EGN Methods for Nonlinearity Assessment of Ultrawideband Coherent Fiber Optic Communication Systems with Raman Effect," *ArXiv Prepr. ArXiv200912687*, Oct. 2020, doi: https://doi.org/10.48550/arXiv.2009.12687.

[15] Y. Jiang, A. Nespola, A. Tanzi, S. Piciaccia, M. R. Zefreh, F. Forghieri, and P. Poggiolini, "Experi-mental Test of Closed-form EGN Model over C+L Bands," 49th European Conference on Optical Communications (ECOC 2023), Hybrid Conference, Glasgow, UK, 2023, pp. 254-257, doi: https://doi.org/10.1049/icp.2023.2053.

[16] Y. Jiang, A. Nespola, A. Tanzi, S. Piciaccia, M. R. Zefreh, F. Forghieri, and P. Poggiolini, "Performance Enhancement of Long-Haul C+L+S Systems by means of CFM-Assisted Optimization," 2024 Optical Fiber Communications Conference and Exhibition (OFC), San Diego, paper M1F.2, CA, USA, 30 Mar. to 03 Apr. 2024.

[17] Y. Jiang and P. Poggiolini, "CFM6, a closed-form NLI EGN model supporting multiband transmission with arbitrary Raman amplification," *ArXiv,* May 2024.

[18] K. Rottwitt, J. Bromage, A. J. Stentz, L. Leng, M. E. Lines, and H. Smith, "Scaling of the Raman gain coefficient: applications to germanosilicate fibers," *J. Light. Technol.*, vol. 21, no. 7, pp. 1652–1662, Jul. 2003, doi: 10.1109/JLT.2003.814386.